\documentstyle[aps,prb,epsfig]{revtex}

\begin{document}
\sloppy
\draft
\title{Phase Diagram of the Spin-Orbital model on the Square Lattice}
\author{E. Zasinas$^{1,2}$, O. P. Sushkov$^1$ and J. Oitmaa$^1$}
\address{$^{1}$School of Physics, The University of New South Wales,\\
Sydney 2052, Australia\\
$^2$Vilnius University, Institute of Materials Science,\\
Sauletekio 9-III, Vilnius, 2040, Lithuania
}
\date{\today}
\maketitle
\begin{abstract}
We study the phase diagram of the spin-orbital model in both the weak
and strong limits of the quartic spin-orbital exchange interaction.
This allows us to study quantum phase transitions in the model and to  
approach from both sides the most interesting intermediate-coupling 
regime and in particular the $SU(4)$-symmetric point of the Hamiltonian. 
It was suggested earlier by Li {\it et al} [Phys.\ Rev.\ Lett. 
{\bf 81}, 3527 (1999)]  that at this point the ground state of the system
is a plaquette spin-orbital liquid. We argue that the state is more
complex. There is plaquette order, but it is anisotropic: bonds in one
direction are stronger than those in the perpendicular direction.
This order is somewhat similar to that found recently in the frustrated
$J_1-J_2$ Heisenberg spin model.
\end{abstract}

\section{Introduction}

In many transition metal oxides the low-lying electron states are
characterized by both spin and orbital degrees of freedom. 
Thus the simplest model that also takes into account the strong Coulomb 
repulsion of transition metal $d$- electrons is the multiband Hubbard 
model. For the case of one electron per site with two-fold orbital
degeneracy (isospin $T=1/2$) and the strong interaction limit Kugel and 
Khomskii \cite{kugel} derived the effective  spin and isospin Hamiltonian:
\begin{equation}
\label{model_hamilton}
H  = \sum_{\langle {\bf i,j} \rangle} 
\left ( \;
J \ {\bf S}_{\bf i} \cdot {\bf S}_{\bf j} \, + \,
J \ {\bf T}_{\bf i} \cdot {\bf T}_{\bf j}  \, + \,
V \ ( {\bf S}_{\bf i} \cdot {\bf S}_{\bf j} ) 
( {\bf T}_{\bf i} \cdot {\bf T}_{\bf j} )
\; \right) \; .
\end{equation}
Here $S_{\bf i}$ and $T_{\bf i}$ are spin and isospin operators
at the lattice site $i$, and the summation is over the nearest
neighbor bonds $\langle {\bf i,j} \rangle$. 
$J$ and $V$ are positive constants of approximately 
equal magnitude. 

In general, the exchange constants in spin and isospin subsystems may differ.
It is also possible that the isospin interaction 
$( T_{\bf i} T_{\bf j} )$ is not spherically symmetric.
If the isospin interaction is spherically symmetric then the total
symmetry of the Hamiltonian (\ref{model_hamilton}) is
$SU(2)\times SU(2)$. At the point $J=V/4$ the Hamiltonian
has even higher $SU(4)$ symmetry. Investigation of the model
at this point has attracted great interest
\cite{firstSU4,su4swave2,exact1d,azaria}. 
In the one-dimensional case there is an exact solution for the ground state
and for several types of excitations \cite{exact1d,firstexact1d}.
For the two-dimensional case (square lattice) an exact solution is not known. 
A kind of mean field $SU(4)$-spin-wave theory has been 
developed for this case \cite{su4swave2,su4swave}, and numerical simulations 
are also available \cite{su4calc,su4calc2}.
The $SU(4)$-spin-wave theory indicates a disordered ground state at $J=V/4.$
However this approach contains uncontrolled approximation, and its
accuracy is uncertain.
It was pointed out in Ref. \cite{su4swave2}  that at $J=V/4$ the
Ising part of the Hamiltonian (\ref{model_hamilton}) is
equivalent to the 4-state Potts model \cite{potts}.
This model has  classical macroscopic degeneracy of the ground state,
and this is another indication in favor of the disordered ground state of
the spin-orbital model at the $SU(4)$ point.
As a possible ground  state the authors of Ref. \cite{su4swave2}
suggested a liquid of plaquette $SU(4)$ singlets.
This ground state is a $SU(4)$ singlet, but it has spontaneous
breaking of the translational lattice symmetry. Further analytical and 
numerical work supports this scenario \cite{su4swave,su4calc,su4calc2}.

The present work has the following goals: to investigate the spin-orbital
model on the square lattice, away from the $J=V/4$ point and to identify all the
quantum transitions, and also to investigate the model away from the
isotropic $SU(2)\times SU(2)$ line. We also present new evidence
in favor of the anisotropic plaquette spin-orbital liquid in the vicinity
of the $SU(4)$ point.
The structure of the paper is as follows. In Sec. II we 
start from the limit $J \gg V/4$ and then approach smaller values
of $J$, using both spin-wave and perturbation methods. 
In section III we consider the opposite limit, of
strong quartic interaction: $J < V/4$. In this case the spin-wave 
theory is not valid and the only tool is a modified series expansion.
Our conclusions are presented in Sec. IV.

\section{The case of weak quartic interaction, $V/4 < J$}

In the limit $J\gg V$ the Hamiltonian (\ref{model_hamilton})
describes two almost independent Heisenberg subsystems.
Both of them are ordered antiferromagnetically, and so the
$SU(2)\times SU(2)$ symmetry is spontaneously broken.
Therefore there are two Goldstone excitations, 
spin-wave ($s$- wave) and isospin-wave ($t$- wave), that determine the
low energy physics of the model. 
The ground state 
is a direct product of the N\'eel-ordered 
spin subsystem and of the N\'eel-ordered 
isospin subsystem, with two sublattices in each subsystem.
In each sublattice we introduce a Dyson - Maleev transformation of spin 
and isospin operators \cite{DM}. For the A sublattice,
\begin{eqnarray}  
\label{eq_dm_UP}
S_{\bf i}^z & = & S- a^{\dagger}_{\bf i} a^{\phantom{\dagger}}_{\bf i}\; , 
\hskip 14 mm 
T_{\bf i}^z  =  T - c^{\dagger}_{\bf i} c^{\phantom{\dagger}}_{\bf i} \; , 
\nonumber \\
S_{\bf i}^- & = & a^{\dagger}_{\bf i} \; , 
\hskip 24 mm 
T_{\bf j}^-  =  c^{\dagger}_{\bf i} \; , 
\\
S_{\bf i}^+ & = & (2S- a^{\dagger}_{\bf i} a^{\phantom{\dagger}}_{\bf i} )
a^{\phantom{\dagger}}_{\bf i} \; , 
\hskip 6 mm 
T_{\bf i}^+  =  (2T- c^{\dagger}_{\bf i} c^{\phantom{\dagger}}_{\bf i} )
c^{\phantom{\dagger}}_{\bf i} \; .
\nonumber
\end{eqnarray}
and for the B sublattice,
\begin{eqnarray}  \label{eq_dm_DOWN}
S_{\bf j}^z & = & -S + b^{\dagger}_{\bf j} b^{\phantom{\dagger}}_{\bf j} \; , 
\hskip 14 mm 
T_{\bf j}^z  =  -T + d^{\dagger}_{\bf j} d^{\phantom{\dagger}}_{\bf j} \; , 
\nonumber \\
S_{\bf j}^- & = & - b^{\phantom{\dagger}}_{\bf j} \; , 
\hskip 24 mm 
T_{\bf j}^-  =  - d^{\phantom{\dagger}}_{\bf j} \; , 
\\
S_{\bf j}^+ & = & -b^{\dagger}_{\bf j} 
( 2S - b^{\dagger}_{\bf j} b^{\phantom{\dagger}}_{\bf j} )\; ,
\hskip 6 mm 
T_{\bf j}^+  =  -d^{\dagger}_{\bf j} 
( 2T - d^{\dagger}_{\bf j} d^{\phantom{\dagger}}_{\bf j} )\; .
\nonumber
\end{eqnarray}
We are interested in $S=T=1/2$, but for now we keep them as
parameters.
After the transformation the Hamiltonian (\ref{model_hamilton})
takes the following form

\begin{eqnarray}
\label{tr_hamilton}
H & = & -{1\over 2} (J - VT^2 )S^2 z N  +
(J - VT^2 ) \sum_{\langle {\bf i,j} \rangle}
\left (
S (
a^{\dagger}_{\bf i} a^{\phantom{\dagger}}_{\bf i} +
b^{\dagger}_{\bf j} b^{\phantom{\dagger}}_{\bf j} -
a^{\phantom{\dagger}}_{\bf i} b^{\phantom{\dagger}}_{\bf j} -
a^{\dagger}_{\bf i} b^{\dagger}_{\bf j}
) + {1\over 2}
a^{\dagger}_{\bf i} 
{ (   b^{\dagger}_{\bf j} - a^{\phantom{\dagger}}_{\bf i}    ) }^2
b^{\phantom{\dagger}}_{\bf j}
\right )
\\
&  & -{1\over 2} (J - VS^2 )T^2 z N  +
(J - VS^2 ) \sum_{\langle {\bf i,j} \rangle}
\left (
T (
c^{\dagger}_{\bf i} c^{\phantom{\dagger}}_{\bf i} +
d^{\dagger}_{\bf j} d^{\phantom{\dagger}}_{\bf j} -
c^{\phantom{\dagger}}_{\bf i} d^{\phantom{\dagger}}_{\bf j} -
c^{\dagger}_{\bf i} d^{\dagger}_{\bf j}
) + {1\over 2}
c^{\dagger}_{\bf i} 
{ (   d^{\dagger}_{\bf j} - c^{\phantom{\dagger}}_{\bf i}    ) }^2
d^{\phantom{\dagger}}_{\bf j}
\right )
\nonumber \\
&  & -{1\over 2} VS^2T^2 z N  +
VST \sum_{\langle {\bf i,j} \rangle}
(
a^{\dagger}_{\bf i} a^{\phantom{\dagger}}_{\bf i} +
b^{\dagger}_{\bf j} b^{\phantom{\dagger}}_{\bf j} -
a^{\phantom{\dagger}}_{\bf i} b^{\phantom{\dagger}}_{\bf j} -
a^{\dagger}_{\bf i} b^{\dagger}_{\bf j}
)
(
c^{\dagger}_{\bf i} c^{\phantom{\dagger}}_{\bf i} +
d^{\dagger}_{\bf j} d^{\phantom{\dagger}}_{\bf j} -
c^{\phantom{\dagger}}_{\bf i} d^{\phantom{\dagger}}_{\bf j} -
c^{\dagger}_{\bf i} d^{\dagger}_{\bf j}
) \; .
\nonumber
\end{eqnarray}
Here $N$ is the number of lattice sites and $z=4$ is the number 
of nearest neighbors. In eq. (\ref{tr_hamilton})
the part containing only the operators
$a$ and $b$ describes the spin subsystem with a renormalized exchange 
constant, $J \to J-VT^2$, while the part containing only the 
$c$ and $d$ operators describes the isospin subsystem with 
similar renormalization, $J \to J-VS^2$.
The other terms describe the interaction between the subsystems.

In the linear spin wave approximation we neglect all quartic terms
in the Hamiltonian (\ref{tr_hamilton}). After this the standard 
\cite{STAND} Bogoliubov transformation diagonalization gives the
dispersion relation for $s$- and $t$-waves:
$\omega_s({\bf k})  =  (J-VT^2)S z\sqrt{ 1 - \gamma^2({\bf k}) }$,
$\omega_t({\bf k})  =  (J-VS^2)T z \sqrt{ 1 - \gamma^2({\bf k}) }$,
where $\gamma({\bf k}) = {1\over 2} (\cos k_x + \cos k_y )$.
The staggered magnetization in this approximation is
independent of $J/V$, and is
\begin{eqnarray}
\label{SzTz}
\langle S_z \rangle &=&S-0.1966,\\
\langle T_z \rangle &=&T-0.1966.\nonumber
\end{eqnarray}

One can also easily take into account single loop corrections to the
linear spin wave approximation. To do this, following the usual
procedure \cite{STAND}, we make all possible decouplings in the 
quartic terms in the Hamiltonian (\ref{tr_hamilton}). This gives
 the following effective Hamiltonian
\begin{eqnarray}
\label{tr4_hamilton}
H & = & -{1\over 2} (J - VT^2 )(S^2 - \kappa^2) z N
-{1\over 2} (J - VS^2 )(T^2 - \kappa^2) z N
-{1\over 2} V(S^2T^2 + 4\kappa^2) z N
\\
& + & 
{\tilde J_s}s \sum_{\langle {\bf i,j} \rangle}
(
a^{\dagger}_{\bf i} a^{\phantom{\dagger}}_{\bf i} +
b^{\dagger}_{\bf j} b^{\phantom{\dagger}}_{\bf j} -
a^{\phantom{\dagger}}_{\bf i} b^{\phantom{\dagger}}_{\bf j} -
a^{\dagger}_{\bf i} b^{\dagger}_{\bf j}
)
\; +\; 
{\tilde J_t} T \sum_{\langle {\bf i,j} \rangle}
(
c^{\dagger}_{\bf i} c^{\phantom{\dagger}}_{\bf i} +
d^{\dagger}_{\bf j} d^{\phantom{\dagger}}_{\bf j} -
c^{\phantom{\dagger}}_{\bf i} d^{\phantom{\dagger}}_{\bf j} -
c^{\dagger}_{\bf i} d^{\dagger}_{\bf j}
) \; .
\nonumber \\
\end{eqnarray}
Here we use the notations:
\begin{eqnarray}
\label{eq_exchangeS}
{\tilde J_s} & = &
(J-VT^2)(1- \kappa) \, +  \, 2V T^2 \kappa,\nonumber\\
{\tilde J_t} &=&
(J-VS^2)(1- \kappa) \, +  \, 2VS^2 \kappa,\\
\kappa & = & 
{1\over S}
\left( 
  \langle a^{\dagger}_{\bf i} a^{\phantom{\dagger}}_{\bf i} \rangle
  -
  \langle a^{\phantom{\dagger}}_{\bf i} b^{\phantom{\dagger}}_{\bf j} \rangle
\right ) 
\, = \,
{1\over S}
\left(
  \langle b^{\dagger}_n b^{\phantom{\dagger}}_{\bf j} \rangle
  -
  \langle a^{\dagger}_{\bf i} b^{\dagger}_{\bf j} \rangle \;
\right ) 
\ = \
{1\over T}
\left(
  \langle c^{\dagger}_{\bf i} c^{\phantom{\dagger}}_{\bf i} \rangle 
  -
  \langle c^{\phantom{\dagger}}_{\bf i} d^{\phantom{\dagger}}_{\bf j} \rangle 
\right ) 
\, = \,
{1\over T}
\left(
  \langle d^{\dagger}_{\bf j} d^{\phantom{\dagger}}_{\bf j} \rangle
  -
  \langle c^{\dagger}_{\bf i} d^{\dagger}_{\bf j} \rangle
\right ) .\nonumber
\end{eqnarray}
The value of $\kappa$ can be calculated in the linear spin-wave 
approximation or it can be found self-consistently.
Both values are very close. For further analysis we take the
self-consistent value: $\kappa= -0.07897/S$. 
The excitation spectra for $s$- and $t$-waves are then
\begin{eqnarray}
\label{exit}
\omega_s({\bf k})& =& {\tilde J_s}Sz 
\sqrt{ 1 - \gamma^2({\bf k}) }, \\
\omega_t({\bf k})& =& {\tilde J_t}T z 
\sqrt{ 1 - \gamma^2({\bf k}) }. \nonumber
\end{eqnarray}
The difference from the linear spin-wave approximation is only
in the renormalized values of $\tilde{J}$.
The staggered magnetization is exactly the same as that in the
linear spin wave approximation, see eq. (\ref{SzTz}).
The  values of the effective exchange constants $\tilde{J}_s$ and
$\tilde{J}_t$ vanish at some critical value of the bare exchange 
constant $J$. One finds from  eqs. (\ref{eq_exchangeS}) that
at $S=T=1/2$ the critical value is
\begin{equation}
J_c = VS^2
{    {S-3\kappa}   \over   {S-\kappa}  } =0.3182V .
\end{equation}
The staggered magnetization given by eqs. (\ref{SzTz}) remains constant
on approaching this point, and therefore the quantum
phase transition at $J/V\approx 0.32$ is of first-order.

In the above analysis  we have taken into account explicitly only the $s$- 
and $t$-waves. However along with these two excitations there are
also the so called spin-isospin excitations ($st$-wave) carrying 
simultaneous spin and isospin flips.
In the spin-wave approach the $st$-excitations are not treated 
explicitly. They appear in the calculation via the quartic
interaction term in the Hamiltonian (\ref{tr_hamilton}) that
we treat in the one-loop approximation.
However as one approaches the transition point this interaction  
becomes more important and our way of treating it is questionable.
Moreover, it is known that in the 1D  model at $J=V/4$ for $S=T=1/2$ 
the $s$- $t$- and $st$-excitations are almost degenerate at $k \to 0$,
see Ref. \cite{exact1d}. This raises further questions about the validity
of our implicit treatment of the $st$-excitation. In the remainder of
this Section we treat the $st$ excitations explicitly.

The spin-wave approach is not convenient for the explicit consideration
of the $st$-excitation. To address this problem we use a kind of 
series expansion method. One takes the Ising part of the
Hamiltonian (\ref{model_hamilton}) as a zeroth-order approximation
and the transverse  terms are treated 
as perturbations proportional to the introduced small parameters 
$x$ and $y$. So we represent the Hamiltonian 
(\ref{model_hamilton}) in the following form
\begin{eqnarray}
\label{series_hamiltonian}
H  & = & H_{0} + H_s + H_t + H_{st} \; ,
\\
H_0 & = &
\sum_{\langle {\bf i,j} \rangle} 
\left ( \,
J S^z_{\bf i} S^z_{\bf j} \, + \,
J T^z_{\bf i} T^z_{\bf j} \, + \,
V S^z_{\bf i} S^z_{\bf j} \ T^z_{\bf i} T^z_{\bf j} 
\, \right) \; ,\nonumber\\
H_s & = & {x\over 2}\sum_{\langle {\bf i,j} \rangle} 
(J+V T^z_{\bf i} T^z_{\bf j} )
( S^+_{\bf i} S^-_{\bf j} \, + \, S^-_{\bf i} S^+_{\bf j} ) \; ,
\nonumber\\
H_t & = & {y\over 2}\sum_{\langle {\bf i,j} \rangle} 
(J+V S^z_{\bf i} S^z_{\bf j} )
( T^+_{\bf i} T^-_{\bf j} \, + \, T^-_{\bf i} T^+_{\bf j} ) \; ,
\nonumber\\
H_{st} & = & { {xy} \over 4} V \sum_{\langle {\bf i,j} \rangle} 
( S^+_{\bf i} S^-_{\bf j} + S^-_{\bf i} S^+_{\bf j} )
( T^+_{\bf i} T^-_{\bf j} + T^-_{\bf i} T^+_{\bf j} ) \, .
\nonumber
\end{eqnarray}
The properties of the initial spherically symmetric Hamiltonian
(\ref{model_hamilton}) are recovered in the  limit $x=y \to 1$.

The effect of the $H_s$ perturbation $(y=0)$ alone leads to the the same
reduction of the staggered magnetization as in the pure Heisenberg 
model.
The average spin is reduced from 1/2 to 
\begin{equation}
\label{sz}
\langle S^z_{\bf i} \rangle = 1/2 -x^2/9 -x^4 4/225 - ... 
\to 0.307.
\end{equation}
The numerical value corresponds to $x=1$, see Ref. \cite{zheng}. 
The antiferomagnetically ordered isospin 
subsystem background results only in a renormalization of exchange 
constant $J\to J-V/4$. However it does not effect the magnetization as 
it does not depend on $J$. 
The same effect occurs with $H_t$ acting alone on the isospin subsystem.

An additional reduction of the staggered magnetization appears due
to the simultaneous action of $H_s$ and $H_t$ ($s+t$ channel). For example
first $H_s$ flips the spins at nearest sites and then $H_t$ flips
the isospins at the same sites. So a virtual fluctuation with
$st$-flips at neighboring sites is created. After that the
fluctuation can collapse back to the Ising ground state using the
same chain of spin and isospin flips. Thus the effect of the
staggered magnetization reduction arises in 4th order of
perturbation theory and hence it is proportional to 
$\sim (J-V/4)^4/J^4$. Near the critical value
 $J \approx J_c \approx 0.32V$,
this reduction is so small that there is no point to calculate it
more accurately.
A more important effect arises from the $H_{st}$ term ($st$-channel)
in the Hamiltonian (\ref{series_hamiltonian}). This perturbation
creates, in first order, a pair of $st$-flips at the neighboring 
sites. The matrix element of the perturbation is $xyV/4$, and the
energy of the virtual excitation is $\Delta E= 6J$. Hence the
reduction of the staggered magnetization due to this mechanism is
\begin{equation}
\label{dds}
\delta \langle S_z \rangle =-4{{(xyV/4)^2}\over{(\Delta E)^2}}=-{1\over{9}}
\left({{xyV/4}\over{J}}\right)^2.
\end{equation}
At the critical value $J\approx J_c \approx 0.32 V$ and at $x=y=1$ 
this gives $\delta \langle S_z \rangle \approx -0.07$. This is the additional 
reduction of the value presented in eq. (\ref{sz}). We see that
the effect of $st$-excitations is not completely negligible, but
it is not qualitatively important: the staggered magnetization
remains finite at the critical point.
This supports our conclusion that the phase transition from the
N\'eel$\times$N\'eel state at $J > J_c \approx 0.32V$
to some other state at $J < J_c$  is of first order.
Unfortunately because of the first-order phase transition this
analysis does not give any insight into the structure of the
ground state at $J < J_c$. A different approach is needed for
consideration of the strong quartic interaction limit.

\section{The strong quartic interaction case, $0<J< V/4$}

We will see that in the limit $J < V/4$ the $st$-excitations are
crucially important. This makes this regime qualitatively different
from that considered in the previous section, and this is why the
spin-wave approach is not helpful at $J < V/4$. We base our analysis
on series expansions using a representation of the Hamiltonian
(\ref{model_hamilton}) in the form (\ref{series_hamiltonian}).
There are two immediately obvious possibilities for the ground state
of the system at small $J$: a) The N\'eel$\times$Ferromagnetic  state 
(N$\times$F), where one subsystem has  N\'eel ordering and the other is
ordered ferromagnetically, b) The Stripe$\times$Stripe state (S$\times$S),
where one subsystem has collinear magnetic order along one crystal axis
and the other subsystem has collinear magnetic order in the perpendicular
direction, see Fig.1. The N$\times$F ground state spontaneously violates
the $SU(2)\times SU(2)$ symmetry of the Hamiltonian (\ref{model_hamilton}).
The S$\times$S state in addition violates the $C_{4v}$ group of the
square lattice, so the total spontaneously broken symmetry in this
case is $SU(2)\times SU(2)\times C_{4v}$.

Now let us look at the Hamiltonian (\ref{series_hamiltonian}). One can
easily show that in the Ising limit (i.e. at $x=y=0$) the  N$\times$F
and S$\times$S states have the same energy. Moreover there is an infinite
set of other degenerate states because a simultaneous spin-isospin flip on 
any lattice site ($st$- wave) does not cost any energy (we consider the
$S=T=1/2$ case). So  in the 
Ising approximation there is an infinite macroscopic degeneracy of the 
ground state. What we want to demonstrate first is that the quantum
fluctuations stabilize the S$\times$S state.

Let us consider the simpleast quantum fluctuations: spin or isospin
flips at nearest sites, $s$- and $t$-waves. These fluctuations are 
generated by $H_s$ and $H_t$ correspondingly, see eq. 
(\ref{series_hamiltonian}).
A straightforward second-order perturbation theory calculation gives
\begin{eqnarray}
\label{NFSS}
E_{N\times F} & = & 
 -\left({V \over 8} + x^2 {{V + 4J} \over 24}\right)N,\\
E_{S\times S} & = & 
 -\left({V \over 8} + (x^2 + y^2)
{  {{(V + 4J)}^2} \over {4 (3V-4J) } }\right)N.\nonumber 
\end{eqnarray}
Thus the $S\times S$ state has lower energy than the $N\times F$ state 
for $x=y$. So, in the following, we consider only the $S\times S$ state.
However eq. (\ref{NFSS}) does not answer the question posed above: what is
the excitation energy of the $st$-wave above the $S\times S$ background?
If it is still zero then the background is unstable.
Analysis of the second-order perturbation theory result shows that the $st$-excitation
blocks some $s$- and $t$-fluctuations, and hence the $st$-excitation energy is
nonzero. A calculation gives the following value of the $st$-gap
\begin{equation}
\label{deffectenergy}
\Delta_{st}=E_{st} - E_{S\times S} =
2(x^2 +y^2)
{    \left(    J+{V\over 4}  \right)    }^2
\left(
{ 4 \over {3V-4J}} - { 1 \over V} - { 1 \over {3V-8J}}
\right)\; , 
\end{equation}
where $E_{st}$ is the energy of the $S\times S$-background with
one $st$-excitation. In this order of perturbation theory the
$st$-excitation is dispersionless.
The gap $\Delta_{st}$ vanishes at $J=0$ and $J=V/4$.
However the gap is positive in the interval $0< J < V/4$ and hence
the quantum fluctuations do, in fact, stabilize the
$S\times S$-state.

One can also calculate the dispersion relations of $s$- and $t$-waves.
To first-order in $H_s$ and $H_t$ the results are
\begin{eqnarray}
\omega_s({\bf k}) & = & V/2 - x(V/4-J)\cos(k_xa) \; , \\
\omega_t({\bf k}) & = & V/2 - y(V/4-J)\cos(k_ya) \; .
\end{eqnarray}
We see that these energies are substantially higher than $\Delta_{st}$.
Therefore below we concentrate on the $st$-waves that drive all the
critical dynamics in the system.

The excitation energy of two remote $st$-waves is $2\Delta_{st}$.
However if two $st$-waves are localized at the nearest diagonal sites of 
the lattice then supression of quantum fluctuations is reduced and a direct
calculation shows that the energy of such a configuration is just
$\Delta_{st}$. This means that two $st$-waves attract each other and
the binding energy is $\epsilon_b=2\Delta_{st}-\Delta_{st}=\Delta_{st}$.
We would like to stress that this is not an effect of a simple potential
attraction, this is the effect of a quantum bag: supression of quantum
fluctuations.

We have found that there is a quantum bag attraction between the
$st$-waves. In this situation it is quite natural to put more $st$-waves
in the bag. Three excitations do not fit naturally into the square lattice,
and therefore we consider four $st$-excitations combined into a plaquette bag
as shown in Fig.2. This is the {\it plaquette} excitation. The corresponding 
excitation energy, the {\it plaquette} gap $\Delta_P$, is equal to
\begin{equation}
\label{plaquettetenergy}
\Delta_P =
2(x^2 +y^2)
{    \left(    J+{V\over 4}  \right)    }^2
\left(
{ 7\over {3V-4J}} - { 4 \over {3V}} - { 1 \over {3V+4J}} - { 2 \over {3V-8J}}
\right)\; .
\end{equation}
The value of $\Delta_P$ versus $4J/V$ is plotted in Fig.3, as the solid
line. For comparison in the same Figure we plot the value of $\Delta_{st}$
given by eq. (\ref{deffectenergy}).
The gap $\Delta_P$ vanishes at $J=0.239V$ and in the interval
$0.239V < J < 0.25V$ it is negative. This means that in this interval the
$S\times S$-state is unstable with respect to condensation of plaquette
excitations. In the phase diagram shown in Fig.4 this region is separated
from that with smaller values of $J$ by the vertical dashed line 
$J\approx 0.24V$.
Naively one would say that a crossing of this line corresponds to a
 second-order phase transition from the $S\times S$-state to
a state with the plaquette order considered in Refs.
 \cite{su4swave2,su4calc}.
However, one needs to examine this more carefully.
As we mentioned above, the $S\times S$-state  violates the
$SU(2)\times SU(2)\times C_{4v}$ symmetry of the Hamiltonian.
On the other hand the plaquette state at $J=V/4$ considered in
Refs. \cite{su4swave2,su4calc} violates
 only $Z_2\times Z_2$-symmetry 
(plaquette state has 2-fold degeneracy in $x$-direction and 2-fold
degeneracy in $y$-direction).
As we have demonstrated above, the $Z_2\times Z_2$ order parameter appears
in the second order phase transition at $J\approx 0.24V$. 
However the question arises: how do the $SU(2)\times SU(2)$
and $C_{4v}$ order parameters disappear when going from small $J$ to
$J=V/4$. To confirm this scenario \cite{su4swave2,su4calc}
we have to find two additional quantum phase transitions:
1) disappearence of $SU(2)\times SU(2)$ order parameters,
2) restoration of $C_{4v}$ order.

The $SU(2)\times SU(2)$ order parameter is the usual staggered magnetization.
So let us calculate reduction of the staggered magnetization.
There are four types of relevant quantum fluctuations that reduce the
magnetization A) spin flips at nearest sites, i.e. a virtual creation of two
$s$-waves, B) isospin flips at nearest sites, i.e. a virtual creation of two
$t$-waves, C) simultaneous spin and isospin flips at nearest sites,
i.e. a virtual creation of two $st$-waves, D) similtaneous spin and isospin
flips at four plaquette sites, i.e. a virtual creation of a {\it plaquette}
excitation. We cannot use a conventional perturbation expansion approach
because, in the Ising approximation,
the excitation energies $\Delta_{st}$ and $\Delta_P$ vanish, see eqs.
(\ref{deffectenergy}) and (\ref{plaquettetenergy}). 
Instead we use the Tamm-Dancoff method \cite{TD} that accounts for all low
energy physical excitations. The average value of $S_z$ is equal to
\begin{equation}
\label{SzTD}
\langle S_z \rangle =
{1\over{2}}-2\left({{t_s}\over{\Delta_s}}\right)^2
-4\left({{t_{st}}\over{\Delta_{st}}}\right)^2
-4\left({{t_{P}}\over{\Delta_{P}}}\right)^2.
\end{equation}
Here $\Delta_s=3V/4-J$ and $\Delta_{st}$, $\Delta_P$ given by eqs.
(\ref{deffectenergy}),(\ref{plaquettetenergy}) are the excitation
energies corresponding to the fluctuations A, C, and D.
The matrix elements $t_s$, $t_{st}$, and $t_P$ are the amplitudes of 
creation of these fluctuations from the Ising ground state.
The fluctuation B (isospin flips at nearest sites) does not contribute to
the reduction of $\langle S_z \rangle$. The coefficients 2, 4, and 4 in 
eq. (\ref{SzTD}) give the number of possibilities for a given virtual 
excitation to contribute to the reduction of $\langle S_z \rangle$.
The amplitude $t_s$ arises in the first order of perturbation theory
in $H_s$, the amplitudes $t_{sp}$ and $t_P$ arise in the second order in 
$H_{st}$ or in the combined fourth order of $H_s$ and $H_t$.
Direct calculation give the following results
\begin{eqnarray}
\label{ttt}
t_s&=& {x\over 2} \left( {V\over 4}+ J \right )\; , \nonumber\\
t_{sp} & =& {{x^2y^2}\over 2} 
{  \left(
               {  { J+{V\over 4} }   \over { {3\over 4}V -J }}
\right )  }^2 
{\left(  J-{V\over 4} \right) }^2
\left( {1\over V} + {2\over {V-2J}} \right) 
+
 {{x^2y^2}\over 4} 
{  \left(
               {  { J+{V\over 4} }   \over { {3\over 4}V -J }}
\right )  }^2 V\; ,\\ 
t_P &=& x^2y^2
{  \left(
               {  { J+{V\over 4} }   \over { {3\over 4}V -J }}
\right )  }^2 
\left( 
{
   {  \left(  J+{V\over 4} \right)^2  }
   \over
   {2V}
}
+
{
   {  \left(  J-{V\over 4} \right)^2  }
   \over
   {V-2J}
}
\right).\nonumber
\end{eqnarray}
Substitution of these amplitudes into eq. (\ref{SzTD}) allows us to find
the reduction of the staggered magnetization. In the final analysis
we set $x=y$. At any given value of $J$
the magnetization vanishes at some particular $x$. This is the location
of the second-order phase transition from the magnetically ordered
phase to the spin-orbital liquid.
In the phase diagram shown in Fig.4 the solid line separates 
these two states.

Thus we have found the transition from the magnetically ordered
$S \times S$-state to the spin-orbital liquid. However we do not see
any mechanism for restoration of the $C_{4v}$-symmetry
violated in the $S\times S$-state.
These considerations lead us to the following phase diagram, see Fig.4.
The phase (I) below the solid line is the Stripe$\times$Stripe phase
which has a nonzero staggered magnetization and also spontaneously 
violates the $C_{4v}$-symmetry of the Hamiltonian.
The phase (II) is the spin-orbital liquid. It has no magnetization,
but still must violate the $C_{4v}$-symmetry. So the symmetry of
this state is exactly the same as the symmetry of the columnar
dimer spin liquid in the frustrated $J_1-J_2$ Heisenberg model at
$0.38 < J_2/J_1 < 0.5$ , see Refs. \cite{Gelfand,Read,Croo,oleg}
Certainly we cannot claim that the state (II) is a kind of 
dimer quantum liquid, but its symmetry is the same.
Finally, the state (III) on the right hand side of the dashed line
is a plaquette spin-orbital liquid because of the condensation
of the plaquette excitations at $J\approx 0.24V$. However it is not
an isotropic plaquette liquid considered in 
Refs. \cite{su4swave2,su4calc}.
According to our symmetry considerations the $C_{4v}$ symmetry
must be violated, so this is a plaquette liquid with bonds in
one direction different from that in the perpendicular direction.
One can say that it is a  Stripe-Plaquette-Correlated
quantum spin liquid. It violates the $Z_2\times Z_2 \times C_{4v}$
symmetry of the Hamiltonian.
This state is similar to that found recently in the frustrated
$J_1-J_2$ Heisenberg model at $0.5 < J_2/J_1 < 0.6$, see 
Refs.\cite{Croo,oleg}.

Our analysis is based on the Tamm-Dancoff expansion at small x.
One certainly cannot guarantee that there is not an additional
phase transition line somewhere at $x\sim 1$. In our opinion
such a qualitative difference is unlikely, but it is possible.
This is why the problem certainly requires further numerical
analysis.

\section{Conclusions}

We have analyzed the properties of the spin-orbital model on the square
lattice in the weak and
strong coupling limits. We show that in the case of weak quartic 
interaction the usual spin-wave approach is valid. There are the 
$st$-quasiparticles that are not adequately described by this approach,
but in this limit these quasiparticles just give small corrections.
In the strong quartic interaction limit the spin-wave approach can
not be applied because all the critical dynamics is driven by the
composite $st$-quasiparticles. The usual series expansion method also
cannot be applied because of the infinite classical degeneracy.
To analyze the situation we apply the Tamm-Dancoff method that proved
to be efficient in this limit.

At $J/V > 0.32$ the ground state of the model is a direct product
of the N\'eel ordered spin subsystem and the N\'eel ordered isospin subsystem.
There is a first order phase transition to the spin-orbital quantum liquid
at the point $J/V \approx 0.32$. We argue that this quantum liquid also 
has a structure.  At $0< J/V < (J/V)_c$ it is a stripe liquid with
spontaneously violated $C_{4v}$-symmetry. At the critical point 
$(J/V)_c$ there is a second-order phase transition to the 
stripe-plaquette-correlated liquid that spontaneously breaks
$Z_2 \times Z_2 \times C_{4v}$ symmetry. The Tamm-Dancoff method gives
the following estimate for the critical point: $(J/V)_c\approx 0.24$.
Thus we argue that the stripe-plaquette-correlated quantum spin-orbital
liquid exists in a narrow interval $0.24 < J/V < 0.32$ around the
$SU(4)$ symmetric point $J/V=0.25$. The structure of this state is
more complex than has been believed previously: it violates an additional
$C_{4v}$-symmetry of the Hamiltonian.

\begin{twocolumn}

\begin{figure} [h]
\centerline{\epsfig{file=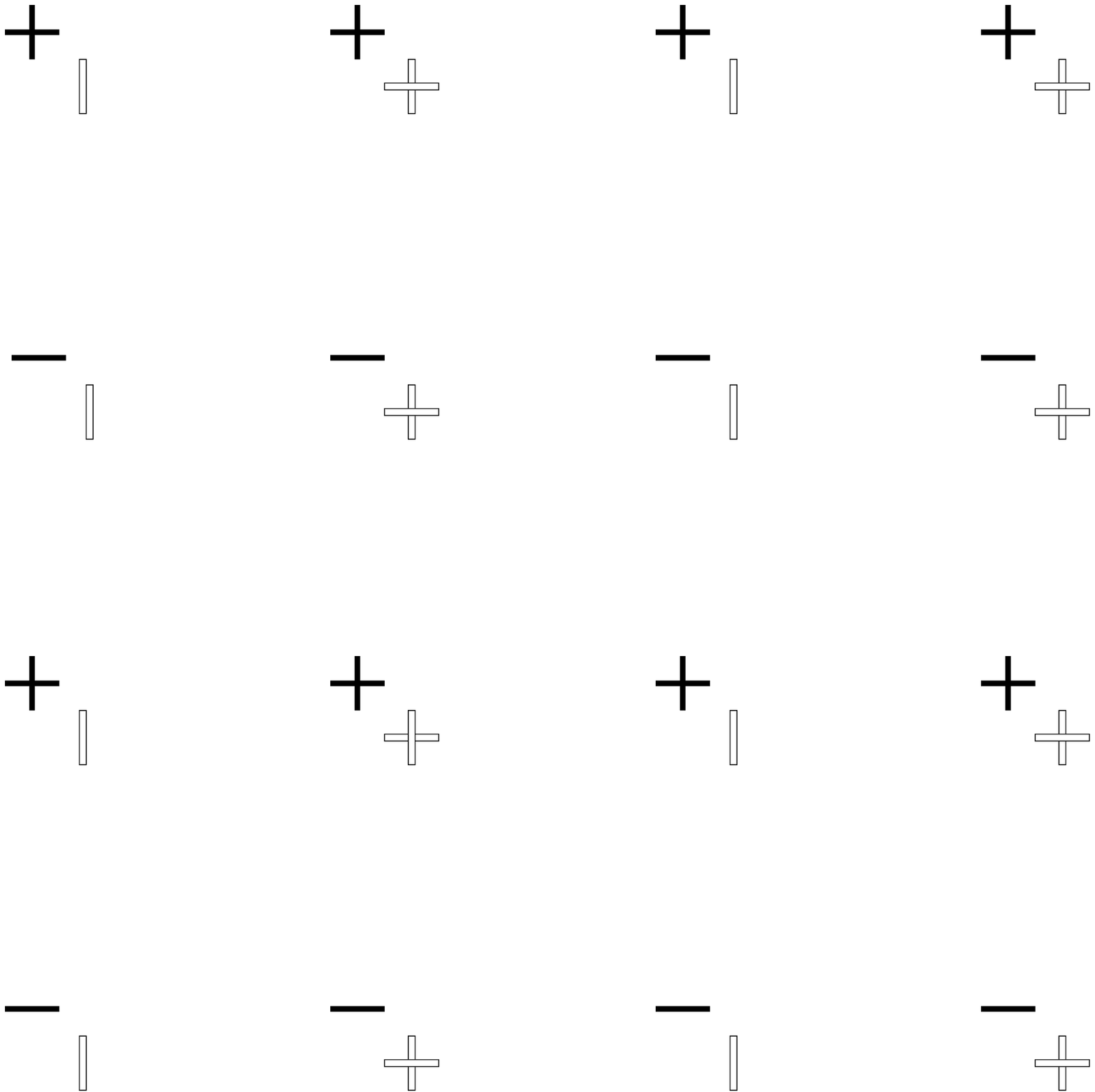, width= 6cm}}
\vskip 5mm
\caption{The structure of the Stripe $\times$ Stripe state.}
\label{fig_stripe}
\end{figure}

\begin{figure} [h]
\centerline{\epsfig{file=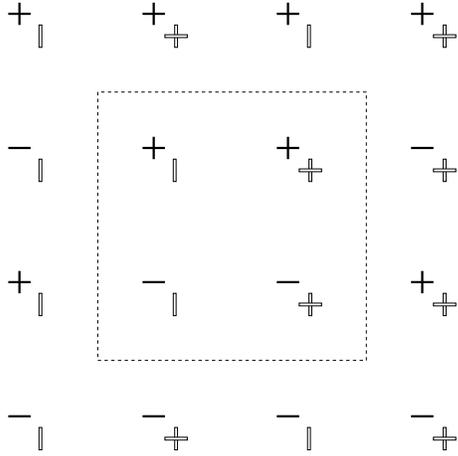, width= 6cm}}
\vskip 5mm
\caption{{\it Plaquette} fluctuation in $S\times S$ background}
\label{fig_plaquette}
\end{figure}

\begin{figure} [h]
\centerline{\epsfig{file=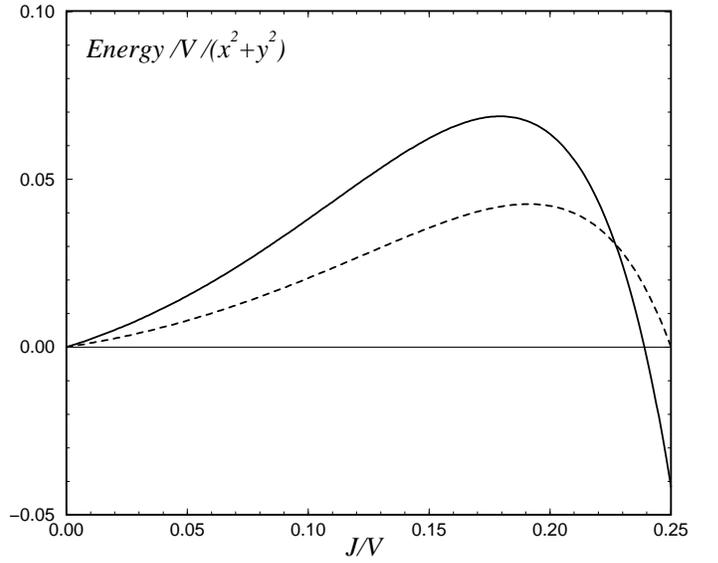, width= 9cm, angle=-90}}
\caption{The plaquette energy gap $\Delta_P$
is shown by solid line. The energy gap $\Delta_{st}$ of the state 
with one $st$-flip is shown by the dashed line.
}
\label{fig_edefect}
\end{figure}

\begin{figure} [t]
\centerline{     \epsfig{file=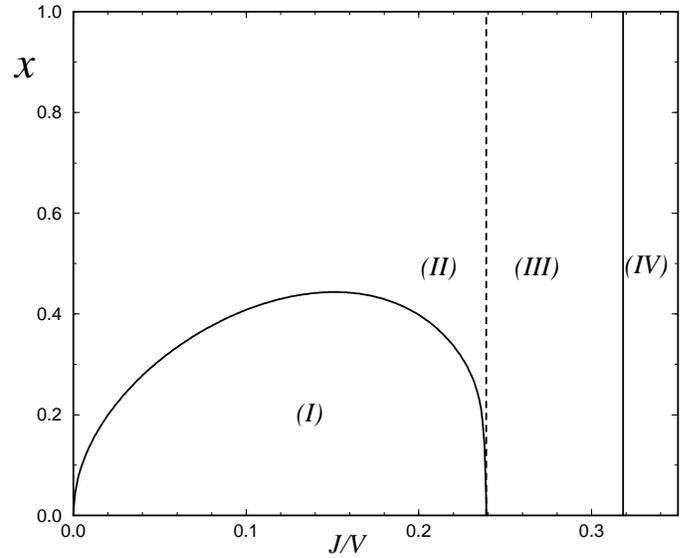, width= 9cm, angle=-90}      }
\caption{
Phase diagram $x-J/V$ of the spin-orbital model on the square lattice.
(I) is the Stripe$\times$Stripe phase with non-zero average spin and 
isospin; (II) spin-orbital liquid; (III) plaquette spin-orbital liquid;
(IV) N\'eel$\times$N\'eel state.
}
\label{fig_phase}
\end{figure}

\end{twocolumn}

\end{document}